\documentclass[letter]{aa} \usepackage{graphicx} \usepackage{epstopdf}
                        \newcommand{\gray}{{$\gamma$-ray}}
\newcommand{\grays}{{$\gamma$-rays}}        \newcommand{\mic}{{$\mu$m}}

\begin{document}

\title{The  Spectrum  of  1ES0229   +  200  and  the  Cosmic  Infrared
Background}

\author{F.W. Stecker\inst{1} \and S.T. Scully\inst{2} }

\offprints{stecker@milkyway.gsfc.nasa.gov}

\institute{NASA Goddard  Space Flight Center, \\  Greenbelt, MD 20771,
USA \and James Madison University, Harrisonberg, VA 22807, USA \\}

\date{Received ; accepted }

\titlerunning{1ES0229 Spectrum and IR Background}

\abstract{}{To   reexamine  the  implications   of  the   recent  HESS
observations   of  the   blazar  1ES0229+200   for   constraining  the
extragalactic  mid-infrared   background  radiation.}{We  examine  the
effect of \gray ~absorption by the extragalactic infrared radiation on
predicted intrinsic  spectra for this  blazar and compare  our results
with  the observational  data.}{We  find agreement  with our  previous
results on  the shape  of the IR  spectral energy  distribution (SED),
contrary  to the  recent assertion  of  the HESS  group. Our  analysis
indicates that 1ES0229+200  has a very hard intrinsic  spectrum with a
spectral index between  1.1 $\pm$ 0.3 and 1.5 $\pm$  0.3 in the energy
range between $\sim$0.5 TeV and $\sim$15 TeV.}  {Under the assumptions
that  (1) the  SED  models of  Stecker,  Malkan \&  Scully (2006)  are
reasonable as derived from  numerous detailed IR observations, and (2)
spectral indexes in the range $1 < \Gamma < 1.5$ have been shown to be
obtainable   from   relativistic    shock   acceleration   under   the
astrophysical conditions  extant in blazar flares  (Stecker, Baring \&
Summerlin 2007), the fits to the observations of 1ES0229+200 using our
previous  IR  SEDs  are  consistent   with  both  the  IR  and  \gray\
observations.   Our  analysis presents  evidence  indicating that  the
energy spectrum  of relativistic particles in  1ES0229+200 is produced
by  relativistic  shock  acceleration,  producing an  intrinsic  \gray
~spectrum with index $1 < \Gamma < 1.5$ and with no evidence of a peak
in the SED up to energies $\sim$ 15 TeV.}

{ }


\maketitle

\section{Introduction}

Shortly after the first strong  \gray ~ blazar 3C279 was discovered by
the EGRET  detector aboard the  Compton Gamma Ray  Observatory (CGRO),
Stecker, De  Jager \&  Salamon (1992) proposed  that the study  of the
spectra  of such  sources could  be  used to  probe the  intergalactic
infrared  radiation.   At  that  time,  there were  no  actual  direct
observations of  the diffuse extragalactic IR  background or extensive
observations of  the sources of such  radiation. The idea  was to look
for  the  effects  of  photon-photon  annihilation  interactions  into
electron-positron pairs. The cross section for this process is exactly
determined; it can be  calculated using quantum electrodynamics (Breit
\&  Wheeler 1934).   Thus, in  principle,  if one  knows the  emission
spectrum  of an  extragalactic source  at  a given  redshift, one  can
determine the  column density  of photons between  the source  and the
Earth.

In the last  15 years, great advances have  been made in extragalactic
infrared astronomy. The diffuse  background at wavelengths not totally
dominated by  galactic or zodiacal  emission has been measured  by the
Cosmic  Background  Explorer (COBE).   In  addition,  there have  been
extensive observations of  infrared emission from galaxies themselves,
whose  total  emission is  thought  to  make  up the  cosmic  infrared
background (see review by Hauser  \& Dwek 2001).  The latest extensive
observations  have been  made by  the Spitzer  satellite.  It  is thus
appropriate  to  use  a  synoptic  approach combining  the  TeV  \gray
~observations with the extragalactic infrared observations in order to
best  explore both  the  TeV  emission from  blazars  and the  diffuse
extragalactic infrared radiation.

Aharonian et  al.  (2007) have recently observed  the spectrum of
the BLLac  object 1ES0229+200 up to  an energy $\sim$~15  TeV with the
High Energy  Spectroscopic System (HESS).  Then, by  assuming that the
intrinsic spectral index of this source is greater than 1.5, they drew
conclusions  regarding wavelength  dependence and  flux of  the mid-IR
extragalactic background  radiation.  Their conclusions  regarding the
mid-IR  extragalactic  background  radiation  appear to  disfavor  the
results   of  the   extensive  semi-empirical   calculations   of  the
extragalactic  IR  background spectrum  given  by  Stecker, Malkan  \&
Scully (2007) (SMS).

In this paper, we  will reexamine the assumptions and conclusions
of Aharonian et  al. (2007) and show that  the observations of 1ES0229
are fully consistent with  the diffuse IR background spectrum obtained
by SMS.  Furthermore, we will show that 1ES0229+200 is an example of a
set of blazars which exhibit  very hard spectra that are indicative of
relativistic shock acceleration.

\section{The Diffuse Extragalactic Infrared Background} 

Various  calculations of  the extragalactic  infrared  background have
been made (Stecker, Puget \& Fazio 1977; Malkan \& Stecker 1998, 2001;
Totani \&  Takeuchi 2002; Kneiske,  et al.  2004; Primack,  Bullock \&
Somerville 2005; Stecker, Malkan  \& Scully 2006, hereafter designated
SMS).  Of these models, the most empirically based are those of Malkan
\& Stecker (1998,2001),  Totanti \& Takeuchi (2002) and  SMS.  The SMS
calculation includes input from the latest Spitzer observations.  
Because the largest uncertainty  in these calculations arises from the
the uncertainty in  the temporal evolution of the  star formation rate
in  galaxies, SMS  assumed  two different  evolution  models, viz.,  a
``baseline''  model  and a  ``fast  evolution''  model.  These  models
produced  similar  wavelength  dependences  for  the  spectral  energy
distribution (SED) of the  extragalactic infrared background, but give
a difference of roughly 30-40\% in overall intensity.

The  empirically  based   calculations  mentioned  above  include  the
observationally based  contributions of  warm dust and  emission bands
from  polycyclic aromatic  hydrocarbon (PAH)  molecules  and silicates
that have been observed to contribute significantly to galaxy emission
in  the mid-IR  (e.g., Lagache  et  al.  2004).   These components  of
galactic infrared emission have the effect of partially filling in the
``valley'' in the mid-IR SED  between the peak from starlight emission
and that  from dust  emission.  In contrast,  the model of  Primack et
al. (2005) exhibits a steep mid-IR valley, which is in direct conflict
with lower limits obtained from galaxy counts at 15 \mic ~ (Altieri et
al. 1999; Elbaz et al. 2002).

\section{Intrinsic Spectra of Blazars from Theoretical Considerations}

The key difference between our  analysis and the analysis of Aharonian
et al. (2007) is their  assumption that the intrinsic spectral indexes
of blazars cannot  be greater than 1.5. Stecker,  Baring and Summerlin
(2007) (SBS) have  shown that spectral  indexes between 1 and  1.5 are
obtainable  from  relativistic shock  acceleration.   They list  three
blazars with intrinsic spectral indexes  between $\sim 1$ and $\sim 1.
5$ in  the energy range between  0.2 TeV and 2  TeV. Their simulations
indicate  that  a  range  of  spectral  indexes  can  be  produced  by
relatvistic  shocks, depending  on  the conditions  in the  individual
shocks.

In fact, recent observations of  an extreme MeV $\gamma$-ray blazar at
a redshift of  $\sim$3 by {\it Swift} (Sambruna et  al.  2006) and the
powerful $\gamma$-ray  quasar PKS 1510-089  at a redshift of  0.361 by
{\it  Swift} and  {\it Suzaku}  (Kataoka et  al. 2007)  both exhibited
power-law spectra in  the hard X-ray range that  had indexes less than
1.5, implying electron  spectra with indexes less than  the value of 2
usually considered for shock acceleration. In addition, the quasar IGR
J22517+2218,  at a  redshift of  3.668,  has been  observed with  {\it
IBIS/INTEGRAL} to have a spectral index of 1.4 $\pm$ 0.6 in the 20-100
keV hard  X-ray energy  range (Bassani et  al.  2007). Spectra  in the
hard X-ray range do not  suffer intergalactic absorption so that there
is no ambiguity concerning their spectral indexes.

\section{The Observed Spectrum of 1ES0229+200 and its Derived Intrinsic 
Spectrum}

The BL  Lac object 1ES0229+200 was  predicted to be  an observable TeV
source  with  a  TeV  flux  of $\sim  10^{-12}$  cm$^{-2}$s$^{-1}$  by
Stecker, de Jager  \& Salamon (1996). It was  detected by Aharonian et
al. (2007) at  a flux level of $\sim 0.94  \times 10^{-12}$ above 0.58
TeV.  Aharonian et al. (2007) have measured the spectrum of the blazar
1ES0229+200 at energies between 0.5 TeV and $\sim$15 TeV.  This blazar
lies at a  redshift of 0.14 (Schachter et  al.  1993).  This redshift,
together with the spectral observations, implies a very hard intrinsic
spectrum for the source.

According to the observations of  Aharonian et al.  (2007), the blazar
1ES0229+200  has  an  observed  spectral  index  of  2.50.   In  their
analysis, Aharonian  et al.  chose  to ``deabsorb'' their  data points
using the results  of various optical depth calculations  and to fit a
simple  power-law spectrum  to the  results.  Using  estimated optical
depths from SMS, they gave  best-fit power-law spectral indexes of 0.1
$\pm$  0.3 for  the fast  evolution  case and  0.6 $\pm$  0.3 for  the
baseline case. We  have done a reanalysis using  their method and find
spectral indexes of  0.2 $\pm$ 0.3 and 0.8 $\pm$ 0.3  for the SMS fast
evolution and  baseline models respectively.  The  small difference in
the results can be attributed  to Aharonian et al.  using interpolated
optical depths,  as SMS  did not provide  the results at  the specific
redshift of 1ES0229+200.  We here  calculated the optical depths for z
= 0.14  exactly, for  both the fast  evolution and  baseline evolution
models of SMS.

In  this paper, we  will adopt  a different  method for  analysing the
intrinsic spectrum  of 1ES0229+200, one  which we feel is  superior to
the above approach.  Aharonian et al.(2007) chose to force a power-law
fit  to  the deabsorbed  data  points. We  choose  here  to assume  an
intrinsic power-law  spectrum emitted by  the source and then  apply a
correction for  optical depth. We then compare  the resulting spectrum
with the actual observed spectrum.  Because of the nonlinear nature of
the energy dependence of the optical  depth, we do not expect that the
observed spectrum will have a power-law form.

Using the  formulas given by Stecker  \& Scully (2006),  we predict an
intrinsic  spectral  index  for  1ES0229+200  of  1.11  for  the  fast
evolution model SED and 1.45 for  the baseline model SED in the 0.2 to
2 TeV  energy range.  This  intrinsic power-law spectrum,  derived for
the 0.2 TeV to  2 TeV energy range, is then assumed  to be extended to
$\sim$15 TeV. Such  a spectrum with an index between 1  and 1.5 can be
obtained  from relativistic  shock acceleration,  as discussed  in the
previous section.  We  have then calculated the form  of the resulting
spectrum  that would be  observed at  Earth by  taking account  of the
absorption predicted by both the baseline and fast evolution models of
SMS.

Thus,   the  observed   spectrum  is   fixed   to  be   of  the   form
$K_{FE}E^{-1.11}e^{-\tau_{FE}         (E;z         =0.14)}$        and
$K_{B}E^{-1.45}e^{-\tau_{B} (E;z  =0.14)}$ for the  fast evolution and
baseline models respectively, where $\tau$ is the optical depth of the
universe  to \grays\  originating  at a  redshift  of 0.14  calculated
according  to SMS.   We employ  a Levenberg-Marquardt  nonlinear least
squares method to fit this form  of the spectrum to the observed data.
In each  case, the  single parameter of  the fit is  the normalization
coefficient, $K$.   The Levenberg-Marquardt method  minimizes $\chi^2$
to best  fit of a nonlinear  function to the  observational data.  For
the     fast    evolution     model,     $K_{FE}    =     10^{-10.80}$
cm$^{-2}$s$^{-1}$TeV$^{-1}$ with  $E$ in TeV; for  the baseline model,
$K_{B} =  10^{-11.13}$ in the  same units.  The resulting  spectra are
plotted along with  the data in Figure 1.  The solid  line is the best
fit  with $\tau$  given for  the baseline  model of  SMS;  the steeper
dashed line is  the best fit with $\tau$ given  for the fast evolution
model of SMS  which gives larger values of $\tau$.  As  can be seen in
Figure   1,  these   curves  show   excellent  consistancy   with  the
observational  data obtained  by the  HESS  group for  1ES0229 +  200.
Indeed, the $\chi^2$ vlaues obtained  are 3.3 for the baseline fit and
5.  7 for the fast evolution fit with 7 degrees of freedom.

  \begin{figure}
   \centering \includegraphics[width=0.45\textwidth]{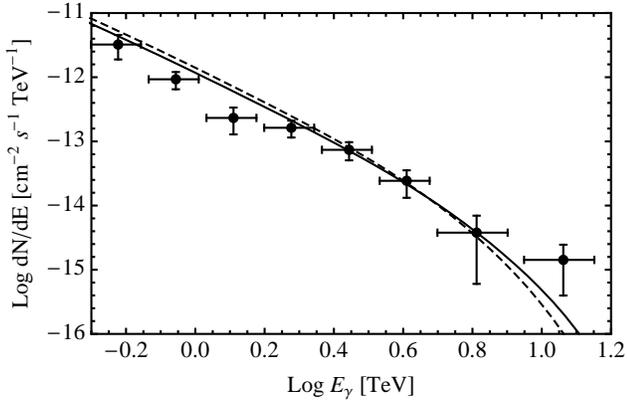}
      \caption{The  observed and  calculated spectrum  of 1ES0029+200.
Data  are  from Aharonian  et  al.   (2007).   The vertical  bars  are
statistical  errors; the  horizontal bars  are the  energy  bins.  The
calculations of the  spectra are as described in  the text.  The solid
line is the best fit with  $\tau$ given for the baseline model of SMS;
the  dashed line  is  the best  fit  with $\tau$  given  for the  fast
evolution model of SMS.}
         \label{}
   \end{figure}

\section{Conclusions and Discussion}

Aharonian et al. (2007), by assuming that blazar spectra have spectral
indexes  $\Gamma >  1.5$, concluded  that the  mid-IR  spectral energy
distribution  (SED) must  have  a wavelength  dependence steeper  than
$\lambda ^{-1}$, in their terminology $\lambda^{-\alpha}, \alpha > 1.1
\pm 0.25$ in the wavelength range  between 2\mic ~ and 10\mic . On the
contrary, galaxy  emission models which take into  account emission by
warm dust  and PAH  and silicate  emission in the  mid-IR, as  well as
direct mid-IR observations of  galaxy spectra give values for $\alpha$
in the range between $\sim$ 0.7  and $\sim$ 0.8 (Spinoglio et al 1995,
Xu  et al.   2001).  These  values  are consistent  with the  2-10\mic
~diffuse background SED given  in the semi-empirical SMS models.  Such
values lead to an energy dependence for the \gray ~optical depth which
is consistent with that obtained by both Totani \& Takeuchi (2002) and
SMS. This  result is also supported  by the lower limit  on the 15\mic
~diffuse background flux obtained by galaxy counts of 3.3 $\pm$ 1.3 nW
m$^{-2}$sr$^{-1}$ (Altieri et al.  1999) which, under the conservative
assumption that 80\% of the  mid-IR flux is resolved out (SMS), yields
a value for the total diffuse background flux at 15\mic ~ of 4.1 $\pm$
1.6  nW m$^{-2}$sr$^{-1}$,  higher  than  the upper  limit  of 3.1  nW
m$^{-2}$sr$^{-1}$  derived  by  Aharonian  et al.   (2007)  under  the
assumption that $\Gamma > 1.5$.

Under the  assumptions that (1) the  SED models of  Stecker, Malkan \&
Scully (2006)  (SMS) are reasonable as derived  from numerous detailed
IR observations, and  (2) spectral indexes in the range  $1 < \Gamma <
1.5$ have been  shown by Stecker, Baring \& Scully  (2007) (SBS) to be
obtainable   from   relativistic    shock   acceleration   under   the
astrophysical conditions extant in blazar flares, the fits to the HESS
TeV  observations  of 1ES0229+200  using  the  SMS  infrared SEDs  are
consistent with both the IR and \gray\ observations.

The  SBS simulations indicate  with specific  test runs  that electron
spectra with asymptotic spectral indexes  between 1.26 and 1.62 can be
obtained from  acceleration by  relativistic shocks with  bulk Lorentz
factors between 10 and 30 and these electrons can then Compton scatter
to produce \gray~ spectra with  indexes $\sim$1.1 and $\sim$ 1.3.  The
simulations  show that  larger bulk  Lorentz factors  lead  to flatter
spectra.  Such  Lorentz factors of 50  or more have  been implied from
studies of specific flares in  BL Lac objects (Konopelko et al.  2003;
Bagelman,  Fabian  \& Rees  2007)  so  that  the existence  of  highly
relativistic shocks in such sources with  indexes as flat as $\sim$1 is 
not unreasonable.

For the SMS fast evolution SED,  the \gray~ index obtained is 1.11 and
for the baseline  SED, the index obtained is  1.45.  Our analysis thus
presents evidence  for relativistic shock  acceleration in 1ES0229+200
that results in a very hard intrinsic \gray ~spectrum with no evidence
of  a peak  in the  \gray ~  SED up  to energies  $\sim$ 10  TeV. 
Unfortunately,  there  are   no  simultaneous  observations  at  other
wavelengths  that can  be used  to model  the flare  that  occurred in
1ES0229+200 and, in any case,  TeV orphan flares have been observed in
other BL Lac  objects.  However, we note that  hard X-ray spectra have
been seen  in other sources as  discussed in Section 3.   We also note
that  SBS have  derived  spectra for  three  other BL  Lac objects  at
redshifts between 0.18 and 0.19  with indexes between 1 and 1.5, viz.,
1ES1218+30, 1ES1101-232, and 1ES0347-121.  It therefore appears that a
whole class of blazars or  blazar flares may exhibit the hard-spectrum
characteristics of relativistic shock acceleration.

\begin{acknowledgements}
We wish to  thank Wystan Benbow for  sending us a list of  data on the
spectrum of 1ES0229+200 observed by HESS. STS gratefully acknowledges
partial support from the Thomas F. \& Kate Miller Jeffress Memorial Trust 
grant no. J-805. We thank an anonymous referee for constructive comments.
\end{acknowledgements}

\end{document}